\documentstyle[preprint,aps]{revtex}

\tighten
\begin{document}
\draft
\preprint{Effect of Cluster Formation on Isospin Asymmetry in the
Liquid-Gas Phase Transition Region}
\title{Effect of Cluster Formation on Isospin Asymmetry in the 
Liquid-Gas Phase Transition Region\\
}
\author{L. Shi and P. Danielewicz\\
}
\address{
National Superconducting Cyclotron Laboratory and
 Department of Physics and Astronomy,\\ 
Michigan State University, East Lansing, MI 48824\\
}
\date{\today}
\maketitle
\begin{abstract}
Abstract: Nuclear matter within the liquid-gas phase transition region is
investigated in a mean-field two-component Fermi-gas model. 
Following largely analytic considerations, it is 
shown that: (1) Due to density dependence of asymmetry energy, some of the 
neutron excess from the high-density phase could be expelled into the 
low-density region. (2) Formation of 
clusters in the gas phase tends to counteract this trend, making the gas phase 
more liquid-like and reducing the asymmetry in the gas phase. Flow of 
asymmetry between the spectator and midrapidity region in reactions
is discussed and a possible 
inversion of the flow direction is indicated.
\end{abstract}

One interesting possibility in heavy-ion collisions at intermediate 
energy is the occurrence of a liquid-gas phase transition. Many 
recent papers addressed this possibility from different perspectives 
\cite{pochodzalla95,D'Agostino99,Chbihi99,Goodman84,Muller95,%
Baldo99,Gupta97}. 
Following an elementary consideration, it is easy to envisage a 
first-order phase transition in infinite nuclear matter. M{\"u}ller 
and Serot first pointed out the importance of isospin for the liquid-gas 
phase transition \cite{Muller95}. 
The additional isospin degree of freedom relaxes 
the system and makes the transition of second-order. Isospin observables 
could generally be used to extract a variety of information from heavy-ion 
collision, see for example the review by Li {\it et al.}  
\cite{Li98,Li95}. Some 
recent data analyses tried to explore isospin observables and to 
relate them to a possible occurrence of the phase transition 
\cite{Dempsey96}. 

One focus of interest in connection with the phase transition 
is the midrapidity region in intermediate-energy heavy-ion collisions
\cite{Dempsey96,Pawlowski98,Toke95}.
In simulations of semiperipheral collisions, a formation of 
low-density neck region is observed that likely contributes to the midrapidity
\cite{Sobotka97}.
 The low-density region in
 contact with high-density regions (the projectile and target) opens up 
the possibility for a liquid-gas phase coexistence and phase conversion.
 In a dynamical simulation with the Boltzmann-Uehling-Uhlenbeck equation,
 Sobotka {\it et al.} \cite{Sobotka97}
 observed neutron enrichment in the low-density neck 
region. However, a high $n/p$ ratio (much higher than in the composite system)
 was found when counting only free nucleons in the neck region, {\it i.e.} 
excluding nucleons in clusters. The paper argued that the symmetric 
clusters (deuterons and alphas) contributed much to the enrichment of
 neutron in the neck region. Specific results were purely numerical 
in nature.

In this paper we shall discuss the isospin asymmetry in the phase 
transition region in a heavy-ion collision and the effect of clusters on 
that asymmetry.
 We will first follow crude statistical arguments, and then construct a 
 model illustrating the same ideas. 

In the general discussion,
let us first allow no cluster formation in an isospin-equilibrated 
 heavy-ion reaction. For a given temperature 
and density, a large isospin asymmetry will increase the total 
energy, which is unfavorable. In a dense phase, the extra energy for
maintaining  the same asymmetry will be much larger than in a dilute phase. 
Thus, if a dilute phase is in isospin equilibrium with a dense phase, 
the asymmetry in the dilute phase will be larger. For the scenario of 
a neck region neighbored by a dense region in heavy ion collision, 
the $n/p$ ratio in the liquid phase is close to that of the whole system, 
while $n/p$ ratio in the gas phase could be much larger than in the 
composite system. Next, we consider letting the clusters be formed 
in the gas phase. Then the available phase space for liquid does not 
change while the phase space for the gas phase increases. The added 
phase space, which corresponds to clusters, has an $n/p$ mean value 
lower than the old phase space for gas. From a statistical equal-partition 
point of view, partition in the new liquid and gas phase space will drive 
the whole gas phase more symmetric. If the percentage of clusters is small, 
however, then there is essentially not much change in the phase space 
distribution, and asymmetry in the gas phase excluding clusters should 
not change much. 

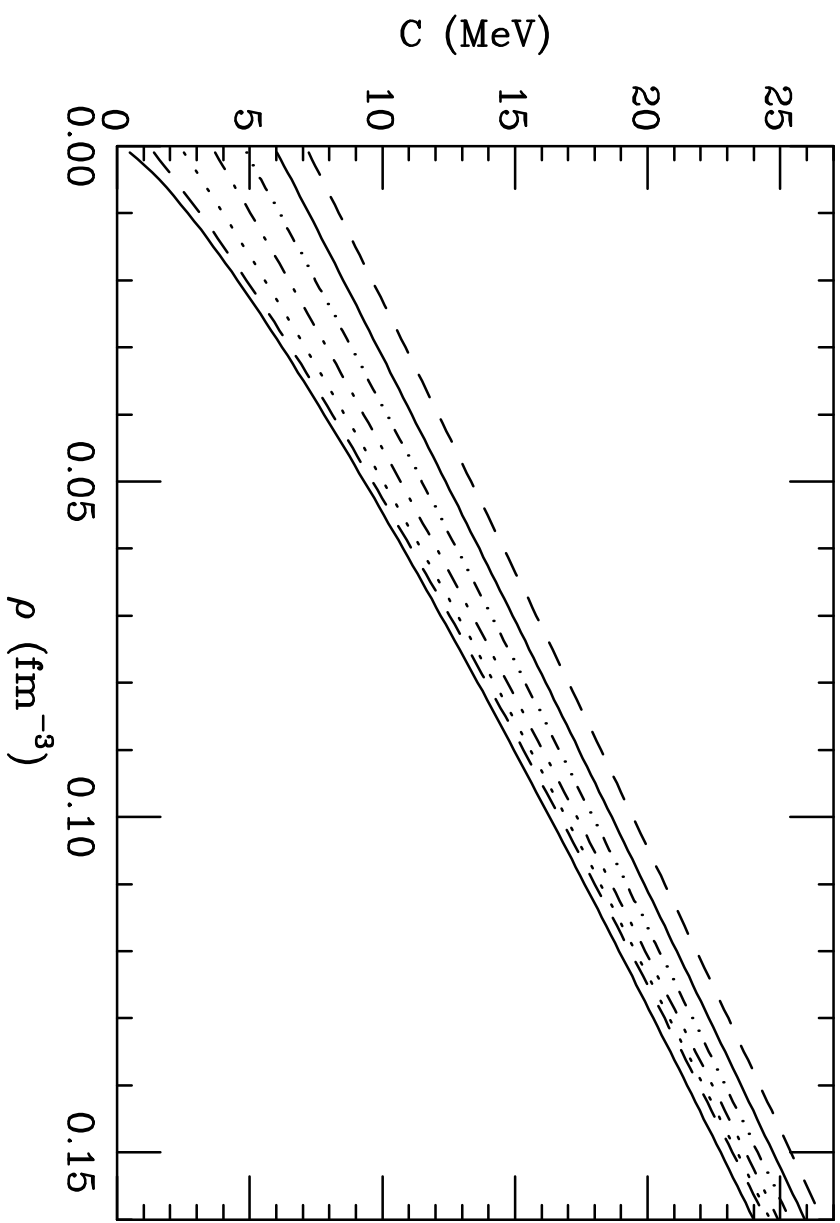
\begin{figure}
\caption{The asymmetry coefficient C as a function of 
density and temperature.
 The lines, from bottom to top, correspond to the temperature of 
0, 2, 4, 6, 8,
10 and 12 MeV, respectively.}
\label{coef-rho}
\end{figure}

Now let us build a simple model and show how isospin equilibrates
 between the two phases. We may start with a two component non-interacting 
Fermi gas of neutrons and protons, and represent the 
interaction by an energy density consistent with the empirical nuclear 
equation of state (EOS). For simplicity we assume no temperature 
dependence for the interaction energy, and the Coulomb interaction 
is not considered here. The total free energy of the system is then 
a sum of the free energies of two non-interacting Fermi gases and 
of a density-dependent nuclear potential energy. 
For a single phase at temperature $T=0$ and density $\rho$, the free energy
 per nucleon may be written as:
\begin{equation}
f={{\cal F} /A} \ =\  
 a_1 \left({{\rho}/{\rho_0}}\right)^{2/3}
+a_2 \left({{\rho}/{\rho_0}}\right)
+a_3 \left({{\rho}/{\rho_0}}\right)^{\sigma-1}
+\left( a_4 \left({{\rho}/{\rho_0}}\right)^{2/3}
 + a_5 \left({{\rho}/{\rho_0}}\right)      \right)y^2
\eqnum{1}\label{eq:1}
\end{equation}  
where $\rho_0$ is the normal density and 
$y$ is the asymmetry parameter, $y=(N-Z)/(N+Z)$.  
For the moment, it is assumed that ${\mid}y{\mid}{\ll}1$. 
The $\left({{\rho}/{\rho_0}}\right)^{2/3}$
 terms come from the non-interacting Fermi gas. The terms
$a_2 \left({{\rho}/{\rho_0}}\right)
+a_3 \left({{\rho}/{\rho_0}}\right)^{\sigma-1}$
 are associated with a simple parameterization 
of the nuclear EOS \cite{Bertsch98,Danielewicz95,Csernai86}. 
As we are only concerned with the isospin 
asymmetry in the liquid-gas phase transition, details of the parameterization 
do not affect our later discussion (though the exact numerical results
may change). Given that the interaction generally contributes to the 
asymmetry energy \cite{Li95}
, we adopt a simple parameterization in Eq.\ (\ref{eq:1}) 
for that contribution, of the 
form $a_5 \left({{\rho}/{\rho_0}}\right)y^2$. 
At $T>0$, the free energy could not be written in a 
simple analytic form, but we can still expand the free energy per 
nucleon about $y=0$. This expansion yields the net 
free energy of the form: 
\begin{equation}
f \ =\ {{\cal F}/A}\ =f_0 +f_y\ =\ f_0 +C y^2  
\eqnum{2}\label{eq:2}
\end{equation}
where $f_0$ and $C$ are functions of both 
temperature and density. The second 
term on the r.h.s. of (\ref{eq:2})
 is due to isospin asymmetry and may be called asymmetric free 
energy. Since our model is symmetric with respect to proton-neutron 
interchange, the expansion of the  free energy contains no odd powers of 
$y$. In our numerical calculation, we use $\rho_0=0.16fm^{-3}$,
$\sigma=2.1612$,
$a_2=-183.05 MeV$, $a_3=144.95 MeV$, $a_5=11.72MeV$,
 and at $T=0$, $a_1=22.10MeV$, and $a_4=12.28 MeV$
($a_4+a_5 \simeq 25MeV$
could be obtained from optical potential analysis \cite{Becchetti69}
 or from the mass formula \cite{Wong98}). 
Numerical analysis indicates that a quadratic form in $y$
 is adequate up to almost 
$y=1$ (a similiar conclusion has been reached in \cite{Prakash88}). 
Figure~\ref{coef-rho} 
 shows the calculated the asymmetry 
coefficient $C$ as a function of density and temperature in the Fermi gas. 
The general 
trend is that $C$ increases with increasing density and temperature.
Therefore, at a given temperature, a dense 
phase will need more extra energy  for
maintaining a given asymmetry than a dilute phase.

Now we can consider a system that has two phases of liquid and gas, 
respectively, in contact with each other. 
The total free energy will be a sum of the
free energies for the two phases. Let us assume that the mechanical 
and thermal equilibrium has been achieved, and now we only consider 
the isospin equilibrium between the two phases. Keeping the total 
asymmetry of the system fixed, we need to vary the asymmetry in 
the liquid and gas phases to minimize total free energy. This yields 
the equilibrium condition:
\begin{equation}
C_l \  y_l \ =\ C_g \  y_g
\eqnum{3}\label{eq:3}
\end{equation}
Here $C_l$ and $C_g$ denote the asymmetric free energy coefficients in the 
liquid phase and in the gas phase. At a given temperature, the liquid 
phase is denser than the gas phase, and the coefficient $C$ is a
 monotonously increasing function of density, $C_l>C_g$. Thus, the 
asymmetry in the gas phase $y_g$ is always larger than that in the liquid 
phase $y_l$. To characterize the relative asymmetry of the two phases, 
we may define the 
isospin asymmetry amplification ratio:
\begin{equation}
 R \ =\ {C_l / C_g} \ =\ {y_g / y_l}
\eqnum{4}\label{eq:4}
\end{equation}

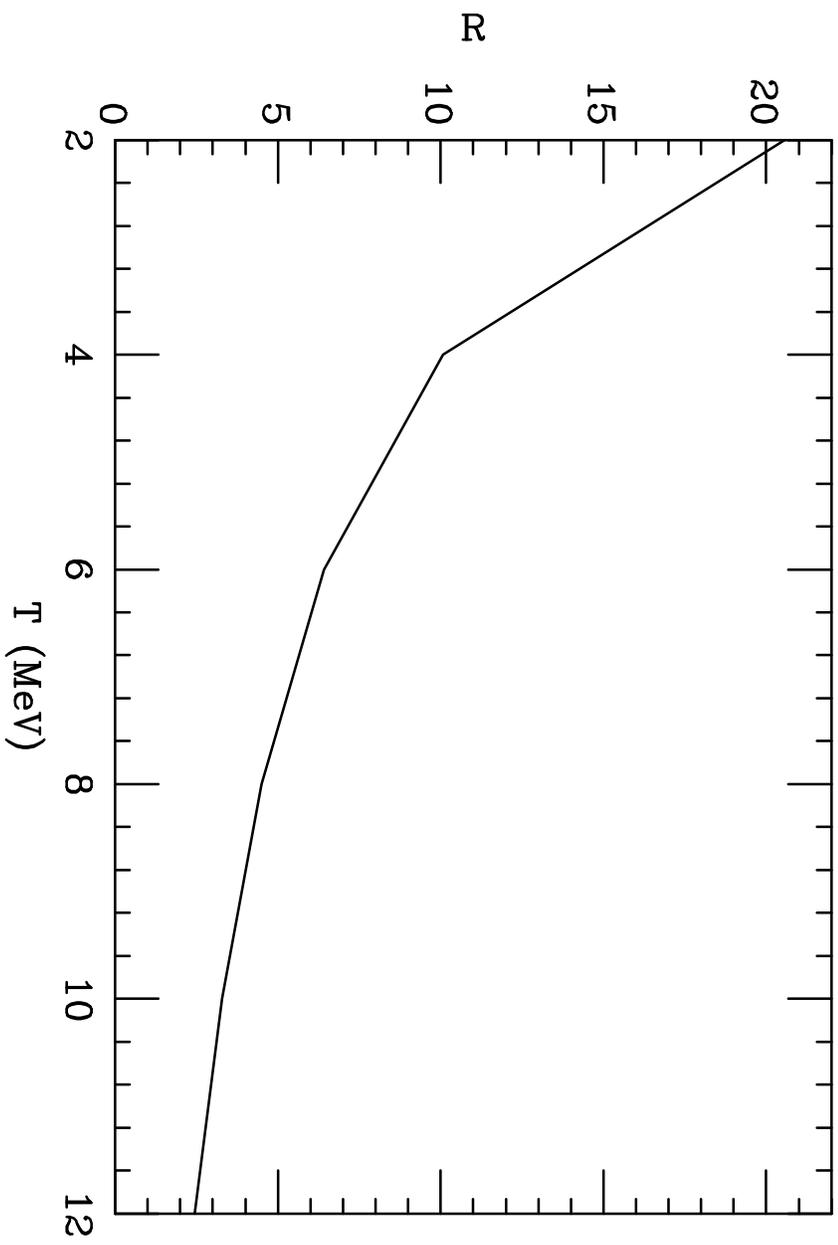
\begin{figure}
\caption{The amplification factor R
 for the
liquid-gas phase transition, as a function of temperature.}
\label{R-T}
\end{figure}

Figure~\ref{R-T} 
displays $R$ vs. temperature for the phases in equilibrium.
For our model calculation, the ratio $R$ stays always larger than $1$, which
 means that the gas phase will always have a higher neutron content
than the liquid phase. Notably the amplification ratio is independent of the 
net isospin asymmetry of the whole system. The ratio $R$ decreases as 
temperature increases, so that a large $n/p$ ratio in the gas phase 
is easier reached at a low temperature.

In the case of a nonequilibrium process,
Eq.\ (\ref{eq:3}) is still of a use due to
a variational origin of the equation. 
If a local equilibrium assumption is met, {\it i.e.}
statistical variables are still valid locally, then 
Eq.\ (\ref{eq:3}) tells us
the direction of developement for the system. The gradient of asymmetry 
coefficient could result in a net flow of isospin asymmetry, which tries to 
restore the isospin equilibrium condition 
Eq.\ (\ref{eq:3}). The flow direction 
is to the steepest decrease of asymmetry coefficient $C$. If there is a 
gradient of density in a nonequilibrium system, hence
a gradient of asymmetry coefficient (see Fig.~\ref{coef-rho}), then there
could be a flow of isospin asymmetry in the system, with the direction
to the low density region. 

We know that, if the nucleon density is not too low, the mean field 
description is quite good. But when the density is low, particle-particle 
correlations become important, and the validity of a mean field 
description worsens. One way to incorporate particle correlations 
is to allow for the formation of clusters in the system (as is done 
in the BUU calculations \cite{Danielewicz91}). 
Since clusters are in practice only important 
for the gas phase, we will only allow clusters there
and no clusters in the liquid phase at all. To further simplify 
the discussion, we shall adopt a droplet model for the clusters (as 
used by Goodman \cite{Goodman84}
 and many others). We will assume that droplets 
have the same properties as the liquid phase, that is the same density 
and asymmetric coefficient; for the present discussion  
we shall ignore the surface 
energy term. Suppose the average size of droplets is $A$, 
and asymmetry in terms of average proton and neutron numbers in droplets
 is $y_d$. The density of nucleons 
in clusters may be represented as $\rho _d=\alpha \rho$ and 
of free nucleons as $\rho _f=(1-\alpha ) \rho$, 
where $\rho$ is the density of the gas phase. The asymmetric 
free energy of the new (free nucleons + droplets) gas phase is:
\begin{equation}
f_y \ =\left(1-{\alpha} \right)\ C_f \ y_f^2 + {\alpha}\ C_d \ y_d^2 .
\eqnum{5}\label{eq:5}
\end{equation}
Here, the subscripts $f$ and $d$ refer to free nucleons and droplets, 
respectively.
To get the isospin equilibrium condition, we can carry out a similar variation 
of asymmetry parameters in the liquid, free-nucleon gas, and in droplets, 
as before, 
obtaining:
\begin{equation}
 y_d \ =\ y_l,\ {\rm and} \ \ {y_f / y_l} \ =\ {C_l / C_f}.
\eqnum{6}\label{eq:6}
\end{equation}
As the density of the gas phase is low,
we may use the ideal gas EOS $p=\rho T$ for clusters in a calculation.
And adding clusters will necessarily decrease $\rho_f$
in order to satisfy the
mechanical equilibrium condition.
However, in Fig.~\ref{coef-rho}  we can see that $C$
decreases only slightly as density decreases. To first order, we can take 
$C_f{\simeq} C_g$, so that $y_f$ is nearly the same as 
the in old gas phase. Overall, 
the asymmetry of the new gas phase is:
\begin{equation}
y \ =\ \alpha \ y_d+\left(1-\alpha \right)y_f.
\eqnum{7}\label{eq:7}
\end{equation}
This may be compared to the asymmetry for the old gas phase, 
$y_g \approx y_f$ , 
which is much larger than $y_d = y_l$. 
It is clear that the more droplets are added to the gas phase, the more it 
looks like the liquid phase. The amplification ratio now is:
\begin{equation}
R\ =\ {y / y_l}\ =\ \alpha+\left(1-\alpha \right) {C_l / C_f}
\approx \alpha+\left(1-\alpha \right) R_0.
\eqnum{8}\label{eq:8}
\end{equation}
where $R_0={C_l/ C_g} \gg 1$. The case of $\alpha=0$
 corresponds to no cluster formation in the 
reaction, and the isospin amplification ratio reaches then the maximum $R_0$. 
The gas phase acquires then the largest possible net asymmetry at a given
temperature. On the other hand, $\alpha=1.0$ corresponds to the gas phase 
with only clusters and the same net asymmetry as for the liquid phase.  

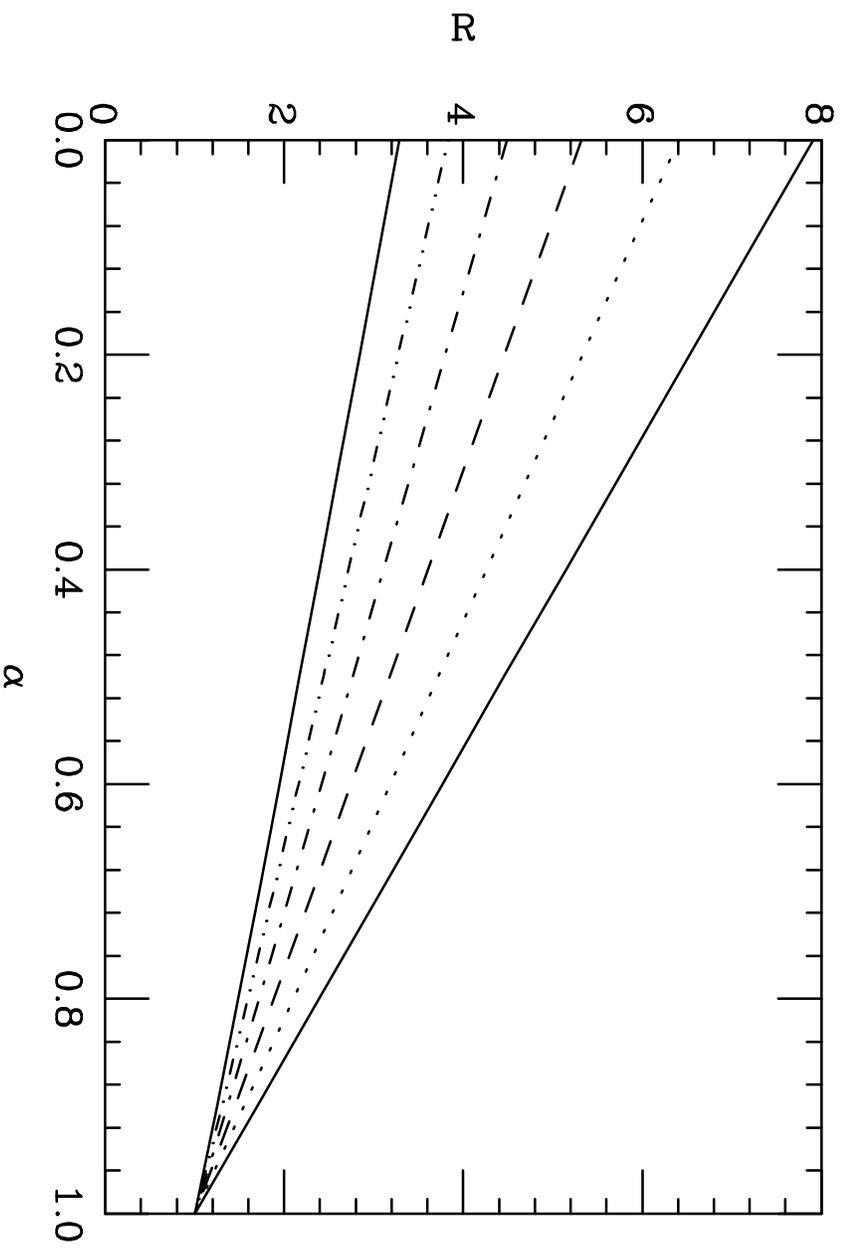
\begin{figure}
\caption{The amplification factor $R$ as a
function of cluster concentration $\alpha$. 
 The lines from top to bottom are for the temperatures of 5,
 6, 7, 8, 9, and 10 MeV, respectively.
}
\label{R-alpha}
\end{figure}

Figure~\ref{R-alpha}  
shows the decrease of the amplification factor $R$ as a function of the 
cluster concentration $\alpha$. 
 As we add more clusters, the low-density gas phase will 
need more energy for the same isospin asymmetry, 
comparable with that of the
liquid phase. As a result, the density and asymmetry in clustered gas 
will both approach 
those in the liquid phase.

Short of simple tools to estimate typical relative numbers of 
free neutrons, free protons and clusters in the gas phase in a reaction, 
we may seek help from experiments. Different regions of velocity 
space are generally believed to reveal characteristics of different sources,
such as the midrapidity particle source for the low-density neck region.
Several intermediate-energy experiments pointed out to a neutron-rich 
midrapidity source in peripheral heavy-ion collisions
\cite{Dempsey96,Toke95,Lukasik97}. Sobotka {\it et al.} \cite{Sobotka00}
measured neutron and $^4He$
emission from a midrapidity source formed
in mid-central $^{129}Xe+^{120}Sn$ collisions 
at 40MeV/nucleon. They compared their
results with results of the INDRA collaboration for the same system 
\cite{Lukasik97,Plagnol99} and gave a
quantitative description of the 
midrapidity source. About half of the charged particles from this source
are $^4He$ and only $10\%$ are free protons.
Similiar results have been obtained in other
papers \cite{Toke95,Lukasik97,Plagnol99,Dempsey96}.
 The number of neutrons is approximately
the same as the number of charged particles, or $10$ times the number of 
protons in this source \cite{Sobotka00}. 
If we take the average cluster size in the midrapidity as about 5 
\cite{cal5}, 
then the percentage
of nucleons inside clusters will be ${\alpha}{\sim}{80\%}$.
The $N/Z$ ratio for the midrapidity source is found to be
higher than for the full system \cite{Sobotka00}. Thus 
the midrapidity source has 
$(N/Z)_{mid}\sim 1.65$ or $y_{mid}\sim 0.25$ while
the system has $(N/Z)_{sys}\sim 1.39$ or $y_{sys}\sim 0.16$. The asymmetry
amplification ratio is then $R\sim 1.5$.
For a mid-rapidity source formed in peripheral heavy-ion collision at similiar
energy, a fully consistent comparision of different experiments is not easy.
Neverthless, comparison of the peripheral data 
from \cite{Dempsey96,Toke95,Larochelle99} also suggests a high cluster 
concentration and a high $n/p$ ratio for free neutron and proton.

In our model calculation, Fig.~\ref{R-alpha}
shows that for the cluster concentration $\alpha$
as high as $80\%$, the asymmetry 
amplification ratio R will decrease by more than a half when compared with
the nonclustered gas phase. This large decrease of R will largely limit
the isospin asymmetry in the gas phase when the asymmetry in the liquid phase
is fixed. Sobotka {\it et al.} 
\cite{Sobotka00} extracted the temperature for the 
midrapidity source
as $6-7$MeV. For this temperature and the cluster concentration
$\alpha\sim 80\%$, we can read off from Fig.~\ref{R-alpha} the 
corresponding equilibrium value as $R\sim (1.9\sim 2.1)$. 
This value is higher 
than the extracted $R\sim 1.5$ in the experiment, which
means that the system only achieved a partial isospin equilibrium and 
the asymmetry amplification in the gas phase did not reach its full value.

While this kind of equilibrium consideration generally give some 
limits for the importance of cluster formation on isospin asymmetry in the 
liquid and gas phase transition region, the developement of
isospin asymmetry
in heavy-ion collision is essentially a nonequlibrium process which 
deserves more thorough investigations than can be comprised in this letter,
possibly incorporating simulations. So we shall
only give some general discussion of the possible isospin asymmetry 
developement in the system.

Because of the transient nature of 
 heavy-ion collision, the development
 of isospin equilibrium could depend on two time scales. One time scale 
is for the separation 
of the midrapidity source from the remaining 
sources, and the other is for isospin 
equilibration. At high enough energy, the three sources 
separate quickly before isospin equilibration could 
set in between sources. The 
isospin asymmetry is then determined by the reaction geometry and the isospin
content
of the target and projectile. Isospin equilibration and cluster formation 
operate only as post-reorganization processes, changing only isospin 
asymmetry for free nucleon and clusters within individual sources.  
The 
large isospin asymmetry for free nucleons could be the result of 
clusterization in the low-density phase, with clusters taking over 
the role of the liquid phase, consistently with 
the arguments by Sobotka {\it et al.} \cite{Sobotka97}.
From our
previous discussion, the R ratio in
Fig.~\ref{R-T} sets an upper limit to
the asymmetry of the free nucleons in the midrapidity source.

On the other hand, if the energy is low enough, partial isospin 
equilibrium will set in before different sources separate from 
each other, and the reaction scenario becomes more complex. As 
the two heavy ions collide against each other, initial compression 
of the participants produces a dense phase in the center, while 
two spectators remain less dense. As the asymmetry 
coefficient for the dense phase is larger than for the
less dense phase (cf. Fig.~\ref{coef-rho}) at the interfaces between the 
two spectator regions and participant region, there could be a local 
density gradient from the center out to the two spectators.
From our arguments following 
Eq.\ (\ref{eq:3}), we know there could then appear an
 isospin asymmetry flow, and it would be out 
to the two spectators. As the compression stage ends, the center 
region begins to expand, and the density drops, the asymmetry 
coefficient also drops as a result. When the gradient of 
the asymmetry coefficient changes direction, the flow 
of isospin asymmetry changes direction too. Cluster formation in the center 
region counteracts the decrease of the asymmetry coefficient, and thus 
delays the change of flow direction. Further development
of the system separates 
the three sources, and net isospin asymmetries for different sources do 
not change after the separation. But clusterization still plays a role
changing the isospin asymmetry of free nucleons within individual sources. 
Since dynamical simulations 
suggest a much longer expansion time than the compression time, we could 
expect that 
the isospin asymmetry flow to the midrapidity region dominates.
This could give rise to an enhanced asymmetry in the midrapidity region. 
The experiments also suggest a neutron-rich midrapidity source, which is 
consistent with the present picture.   

In conclusion, we have investigated the isospin asymmetry in the nuclear 
liquid-gas phase-transition region. In the framework of the 
two-component Fermi-gas with a parameterized 
interaction, under the assumption of isospin equilibrium, we
found that a neutron enrichment in the gas phase is due to the 
density-dependent part of the asymmetry energy. 
Meeting the isospin equilibrium condition, 
Eq.\ (\ref{eq:3}), 
drives extra neutrons out to the low-density phase.  
The formation of 
clusters, which have average asymmetry smaller than the gas phase,
will make the gas phase more liquid-like, and counteract the
neutron enrichment in the gas phase. 
The $^4He$ clusters will be the most important due to their predominance in the 
neck region \cite{Toke95,Dempsey96}. Based on the isospin 
equilibrium requirement, isospin asymmetry flow was suggested if there
exists a local density gradient in heavy-ion
collisions. Since the midrapidity undergoes compression and expansion, we 
also suggested a possible change of the direction of the isospin asymmetry 
flow during the evolution of the system.

\begin{center}
***
\end{center}

This work was partially supported by the National Science 
Foundation under Grant PHY-9605207.

\begin{references}
\bibitem{cal5} For an estimate of the average size of clusters in the 
midrapidity source, we can take data from 
ref. \cite{Lukasik97}, and make a simple calculation.
\end{references}
\end{document}